\documentclass[graybox]{svmult}

\usepackage{mathptmx}       
\usepackage{helvet}         
\usepackage{courier}        
\usepackage{type1cm}        
\usepackage{makeidx}         
\usepackage{graphicx}        
\usepackage{multicol}        
\usepackage[bottom]{footmisc}

\makeindex             

\begin{document}

\title*{Theories of the massive star formation: a (short) review}
\author{Patrick Hennebelle and Beno{\^ i}t Commer{\c c}on}
\institute{Patrick Hennebelle \at Ecole normale sup\'erieure
 \email{patrick.hennebelle@ens.fr}
\and Benoit Commer{\c c}on \at Ecole normale sup\'erieure
 \email{benoit.commercon@ens.fr}}
%
%
\maketitle

\abstract{We briefly review the recent numerical works that have been 
performed to understand the formation of massive stars.
After a brief description of  the classical works, we review
more specifically $i)$ the problem of building stars more 
massive than 20 $M _\odot$ and $ii)$ how to prevent the massive 
cores to fragment in many objects.  Multi-D 
simulations succeed in circumventing the radiative pressure leading 
to the formation of massive stars although 
some questions are still debated
regarding how is accretion exactly proceeding. 
While the core fragmentation  is
slightly reduced by the  radiative feedback and the magnetic field when they 
are treated separately, it is almost entirely suppressed when 
both of them are included. This is because, magnetic field by removing 
angular momentum focusses the flow in a compact region. This makes
the radiative feedback very efficient leading to a significant
increase of the temperature.}

\section{Introduction}

High-mass stars have stellar masses roughly 
spanning the range $10-100~M _\odot$. From their birth to their death, 
high-mass stars are known to play a major role in the energy budget of 
galaxies via their radiation, their wind, and the supernovae. Despite
 that, the formation of high-mass stars remains an enigmatic process, 
far less understood than that of their low-mass (solar-type) counterparts. 
 One of the main differences between the formation 
of  high-mass and 
low-mass stars is that the radiation field of a massive protostar plays a 
more important role. Indeed, the massive stellar embryo strongly heats the
 gas and could even prevent further matter accretion through its radiation
 pressure. This  implies that the
 radiative transfer must be treated in parallel to the hydrodynamics, 
which represents a severe complication mainly responsible for the limited 
numbers of theoretical studies of this process. As described
below, it has been realised that the magnetic field is 
likely to play an important role  as well.

Here is presented a short introduction to the theory of high-mass star 
formation.   
We first present the basic principles used 
to estimate the largest stellar mass that one expects to form in the 
presence of  radiative forces in 1D. We then describe 
the recent numerical simulations  which have been performed 
to address the two important questions; how to build 
stars more massive that 20 $M_\odot$ in spite of the 
radiative pressure; and how to prevent massive cores
from fragmenting in many low mass objects ?

\section{The issue of circumventing the radiative pressure}

\subsection{One dimensional estimate}

The first estimates of the largest stellar mass that can possibly be
 assembled are due to Larson \& Starrfield (1971) 
and Kahn (1974). The principle of their analysis is to compare 
the radiative pressure of a massive stellar embryo to the ram pressure
 induced by the gravitational collapse of its surrounding massive cloud, 
in its inner and outer parts. If the luminosity of the central star becomes
 high enough, the radiation pressure may become important and prevent 
further accretion onto the central object. Since the radiation pressure is
 acting on the dust grains, one has to assume that the frictional coupling
 between the gas and the dust is sufficiently strong so that forces acting
 on the dust grains are transmitted to the gas. 

In the inner part of the collapsing cloud, the temperature becomes high and the dust grains evaporate. There is thus a dust shell whose inner edge is located at the radius, $r$, where the grains evaporate. At this sublimation
radius, the radiation pressure is $L_\star / 4 \pi r^2 c$, where $L_\star$ is the stellar luminosity and $c$ the speed of light. The dynamical pressure is $\rho u^2$, where $\rho$ is the density and
$u$ the infall speed which is given by $u^2 \simeq 2 G M_\star / r$, where $G$ is the gravitational constant and $M_\star$ the mass of the protostar. This leads to the ratio of radiative to ram pressures
	\begin{eqnarray}
	\Gamma  = { L_\star / 4 \pi r^2 c \over \rho u ^2}
	\simeq 1.3 \times 10^{-11}~ { L_\star / L _\odot \over (M_\star / M _\odot) ^{1/2}} ~r^{1/2}. 
 	\end{eqnarray}
Using an analytic estimate for the temperature inside the cloud and based on the assumption that the grains 
evaporate at a temperature of $\sim$1\,500~K, Larson \& Starrfield (1971) estimate the radius of the 
shell to be
	\begin{eqnarray}
	r \simeq 2.4 \times 10^{12} {(L_\star / L _\odot) ^{1/2} \over (M_\star/M _\odot)^{1/5}} \, {\rm cm} 
	\simeq 3.3~ {(L_\star /10^3~ L _\odot) ^{1/2} \over (M_\star/8~M _\odot)^{1/5}} \, {\rm AU}.
	\end{eqnarray}
It follows from Eqs.~2.1--2.2 that 
	\begin{eqnarray}
	\Gamma \simeq 2 \times 10^{-5}~ {(L / L _\odot) ^{6/5} \over (M/M _\odot)^{3/5}}.
	\end{eqnarray}
For a stellar mass of $20~M _\odot$, corresponding to a luminosity of about $4 \times 10^4~L _\odot$, $\Gamma$ roughly equals unity. Therefore, according to Larson \& Starrfield (1971), the mass at which radiative pressure impedes accretion is around $20~M _\odot$. 

 A more accurate estimate has been done by Wolfire \& Cassinelli (1987) by using the optical properties and composition of the mixture of dust grains proposed by Mathis et al. (1977). Assuming an accretion rate of $10^{-3}~M _\odot\,$yr$^{-1}$ in a $100~M _\odot$ cloud, Wolfire \& Cassinelli show that $\Gamma$ is larger than one for any reasonable value of the radiation temperature. They conclude that building a massive star with the ``standard'' dust grain mixture
 is difficult and requires reducing the grain abundance by large factors ($\sim$4--8). They thus propose, as a solution to the high-mass star formation problem, that the dust abundance could be locally decreased by an external shock or an internal ionization front.

More recently, Kuiper et al. (2010) have also performed 1D calculations
for various core masses and confirm largely the results of these
early works. In particular, they cannot form objects more 
massive than 20 $M_\odot$ even in very massive cores.


\subsection{Bidimensional multi-wavelengths calculations}

Bi-dimensional numerical simulations have been performed, 
treating the radiation and the dynamics self-consistently. 
In these studies, it has been assumed that the radiation arises from both
 the accretion and the stellar luminosity. While the former is dominant
 during the earliest phases of the collapse, the latter becomes more 
important at more advanced stages. 
 One of the main motivations of these calculations is to determine whether
 the presence of a centrifugally supported optically thick disk, inside 
which the radiative pressure would be much reduced, may allow to circumvent
 the radiation pressure problem. 
The first  numerical simulations have been performed by
 Yorke \& Sonnhalter (2002) in the frequency dependent case 
(using 64 intervals of frequency) and in the grey case (one single interval
 of frequency). The cloud they consider is centrally peaked, has a mass 
of 60 $M _\odot$, a  thermal
 over gravitational energy ratio of about 5$\%$ initially, and is slowly 
rotating. 
 After $\sim$10$^5$~yr, the central core has a
 mass of about $13.4~M _\odot$ and the surrounding cloud remains nearly
 spherical. After $\sim$2$\times 10^5$~yr, the mass of the central core 
is about $28.4~M _\odot$ and the cloud starts to depart from the spherical
 symmetry. In particular, the infall is reversed by radiative forces in the
 polar region while the star continues to accrete material through the 
equator where the opacity is much higher. This is known as the 
  ``flashlight effect''.  
Once  the stellar mass has grown to about $33.6~M _\odot$, the central star 
is no longer accreting although $30~M _\odot$ of gas is still available 
within the computational grid. The infall is then reversed in every 
directions indicating that the radiative forces are effectively 
preventing further accretion. 
If instead of a multi-frequency treatment, the grey approximation is
 made, the early evolution is similar but becomes notably different after
 $\sim$2.5$\times 10^5$~yr. In particular, there is no evidence of any 
flow reversal. Instead the material flows along a thin disklike structure,
 supported in the radial direction by both centrifugal and radiative forces.
 At the end of the simulation, the mass of the central star is about 
$20.7~M _\odot$.

Kuiper et al. (2010) have performed bi-dimensional simulations 
using an hydrid scheme for the radiative transfer. While the 
gas emission is treated using the flux-limited diffusion and the 
grey approximation, direct multi-frequency irradiation from the 
central star is also included. 
In particular, they stress the importance of spatially resolving the 
dust sublimation front.
In the simulations that do not resolve it well,
the accretion quickly stops while it continues when the sublimation 
front is well described. This is because the radiation is more 
isotropic when the dust sublimation front is not properly 
resolved, leading to a 
weak flashlight effect. In their simulations, Kuiper et al. (2010)
form objects of mass much larger than  the $\simeq$20$M_ \odot$ that
they 
form in their 1D calculations. For example for a 480 $M _\odot$ clump, 
they form an object of 150 $M _\odot$ which is still accreting.

\subsection{Tridimensional calculations}

The first 3D-calculations have been performed by Krumholz et al. (2007, 2009).
They use a flux-limited and grey  approximation to treat 
the radiative transfer. The most striking aspect they report 
is certainly the development of the Rayleigh-Taylor 
instability in the radiatively triggered expanding bubble. 
As a consequence of the non-linear development of this 
instability, fingers of dense material can channel through 
the low density radiatively dominated cavity and reach 
the central object. They therefore identify three modes 
of accretion in their simulations, accretion through 
the disk (the flashlight effect), accretion through the cavity 
wall, and accretion through dense Rayleigh-Taylor unstable fingers.
A quantitative estimate reveals that the latter route accounts for 
about 40$\%$ of the accretion. 

These results have been questionned by Kuiper et al. (2012) 
who performed bidimensional calculations with a flux-limited
scheme similar to the one used by Krumholz et al. (2009) 
and the hybrid scheme which is used in Kuiper et al. (2010).
The results turn out to be quite different.
In the first case, a radiatively dominated bubble is launched
but is quickly stopped and falls back towards the equatorial 
plane. In the second case, the bubble keeps expanding leading 
to a radiatively driven outflow. One of the important consequence
is thus that accretion occurs exclusively through the disk. 
As these simulations are 
bidimensional, it is unclear whether they completely rule out 
the development of the Rayleigh-Taylor instability which 
could be largely seeded by the non-linear fluctuations 
induced by the turbulence in 3D. They nevertheless
suggest that the dynamics of the radiatively dominated cavity
is largely determined by 
the treatment of the radiative feedback in particular its 
frequency dependence.

\section{The issue of fragmentation}

The second drastic problem in the context of massive star formation 
is how to avoid fragmenting the massive cores in many objects. 
For example in the simulations that have been performed by 
Dobbs et al. (2005), the 30 solar mass core they simulate, fragments
in about 20 low mass objects thus preventing the formation of 
high mass objects. While it remains possible that large mass objects
could be formed in very massive clumps through competitive  
accretion (e.g. Bonnell et al. 2004), 
it is important to treat in any case, the physics
of the fragmenting cores properly which is the task that the studies
described below have addressed. 

\subsection{Hydrodynamical radiative calculations}

\begin{figure}[t]
\includegraphics[scale=0.6]{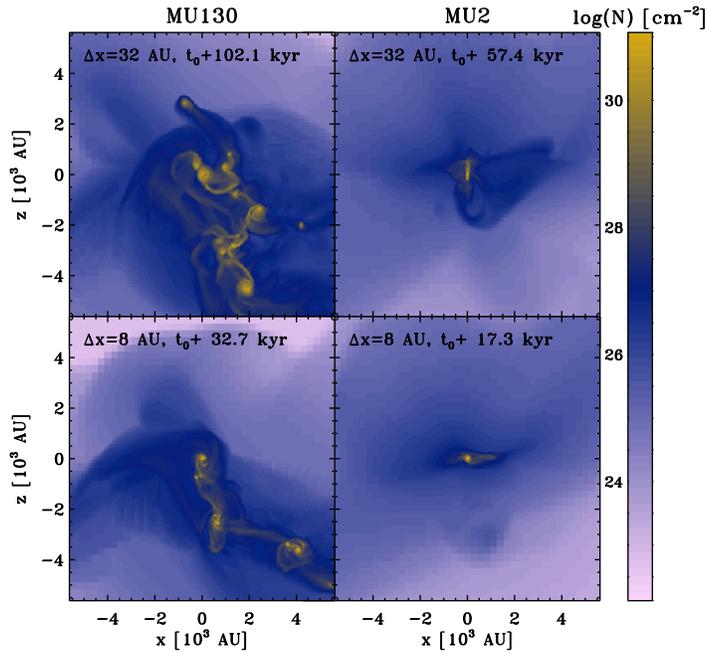}
\caption{Column density within central part of massive 
collapsing cores. Left: hydrodynamical case, the 
core is largely fragmenting eventhough radiative feedback is treated.
Right: RMHD simulation (initial mass-to-flux of 2), the fragmentation 
is entirely suppressed due to the combination of magnetic field 
and radiative feedback (Commer{\c c}on et al. 2011). }
\label{col_dens}
\end{figure}

Tridimensional calculations have been performed by
 Krumholz et al. (2007) using the grey approximation for the 
radiative transfer. Their initial conditions (aimed at reproducing the model
 of McKee \& Tan (2003) consist in a  centrally peaked $100~M _\odot$
 cloud with a density profile proportional to $r^{-2}$. The initial 
turbulence within the cloud is sufficient to ensure an approximate 
hydrostatic equilibrium. Turbulent motions first delay the onset of 
collapse but, as the turbulence decays, the cloud starts to collapse.
 Comparison is made with runs for which an isothermal equation of state
 is used.
In particular, Krumholz et al. (2007) find that, when the
 radiative transfer is taken into account, the gas temperature inside the
 cloud is higher than in the isothermal case, by factors up to 10, which 
are depending on the cloud density.  
 As a consequence, the cloud is fragmenting much less when 
 radiation is taken into account than when isothermallity
 is used. 
It is important to note at this stage that  centrally 
condensed cores are less prone to fragmentation than 
cores having flatter density profiles as shown 
by Girichidis et al. (2010). Indeed the 
radiative hydrodynamical 
simulations performed by Commer{\c c}on et al. (2011) 
clearly show that cores which  initially
have a flat density profile, are undergoing 
significant fragmentation as shown by the top left
and bottom left panels of Fig.~\ref{col_dens}.

\subsection{MHD barotropic calculations}
Another important process that must be included 
in the treatment of massive cores, is the magnetic 
field. Indeed, in the context of low mass cores, 
magnetic field has been found to drastically 
reduce the fragmentation 
(Machida et al. 2005, Hennebelle \& Teyssier 2008).
Hennebelle et al. (2011) have been running a set 
of barotropic simulations for various magnetic intensities. The 
initial conditions consist in 100 $M _\odot$ cores
with a smooth initial density profile and a turbulent 
velocity field (with a ratio of turbulent and gravitational
energies of about 20$\%$). 
The fragmentation is delayed and reduced when 
the magnetic flux is strong enough (typically 
for mass-to-flux smaller than 5). 
The number of objects decreases up to typically 
only a factor  of two for the strongest magnetisation that 
was explored. Thus, Hennebelle et al. (2011) conclude
that magnetic field in itself cannot suppress the fragmentation
in many objects. The reason of this limited impact 
is largely due to the magnetic diffusion induced by the
turbulent velocity field, which reduces the magnetic field
in the central part of the collapsing core where 
fragmentation is taking place.
Similar conclusion has been reached by Peters et al. (2010) 
who even included photo-ionisation from the central star.

\subsection{MHD radiative calculations}

\begin{figure}[t]
\includegraphics[scale=0.4]{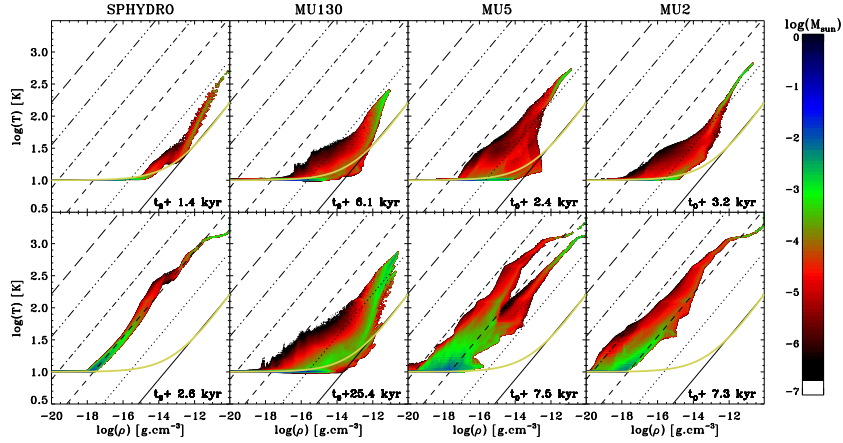}
\caption{Temperature as a function of density within 
the massive core. First panel is for a purely spherical 
model (i.e. which has no turbulence and no magnetic field), 
the second panel is for an hydrodynamical run with initial
turbulence, the third is identical to the second one but 
has an initial magnetic field corresponding to 
mass-to-flux of 5 initially. The fourth has a mass-to-flux of 2
(Commer{\c c}on et al. 2011).}
\label{temperature}
\end{figure}

The first simulations that include both MHD and radiative feedback
in the context of massive star formation, have been recently 
performed by Commer{\c c}on et al. (2011). These simulations
show that the combination of magnetic field and radiative feedback 
is indeed extremely efficient in suppressing the fragmentation. 
The reason is that magnetic field and radiative feedback 
are in a sense interacting (Commer{\c c}on et al. 2010) 
and their combination leads 
to effects that are much stronger than expected. This is 
because, as pointed out by Hennebelle et al. (2011) 
magnetic field, even in the presence of turbulence, leads 
to efficient magnetic braking which reduces the 
amount of angular momentum in the central part of the cloud 
where fragmentation is taking place. Thus, the accretion 
is initially much more focussed in a magnetized core 
than in an hydrodynamical core when turbulence is included
because in hydrodynamical simulations, a large amount 
of angular momentum prevents the gas to fall in the central
object. Consequently, the accretion luminosity which 
is $\propto M \dot{M} / R$ is much higher because the mass
of the central object and the accretion rate onto the central 
object are larger. Also the radius at which accretion is stopping 
is smaller (since there is less angular momentum).
Consequently, the temperature in magnetized cores is much higher 
than in hydrodynamical cores making them much more stable 
against fragmentation. This is illustrated in Fig.~\ref{temperature}
which shows the temperature as a function of density
in four cases. The first panel shows the case of a cloud
with no turbulence and no magnetic field which is purelly 
spherical initially. In this case, the flow is extremely 
focussed and fall directly in a single central object. 
The second panel shows the temperature
distribution for a turbulent and unmagnetized cloud while the 
third and fourth panels show this distribution for two
 magnetic intensities. Clearly the hydrodynamical case 
with turbulence has the lowest temperatures while the 
most magnetized case (fourth panel) presents much 
higher temperatures which are comparable to the one 
obtained in the purely 
spherical case (first panel) that is naturally focussed.

\section{Conclusion}
We have presented a brief review of the recent studies which have been 
performed to explain the formation of massive stars. 
Multi-D simulations including radiative feedback agree
that it is possible to build stars more massive than 
predicted by the 1D spherical case in which radiative 
pressure prevents further accretion. The details, however,
of how this accretion exactly proceeds are still a
matter of debate. The fragmentation of massive cores
is slightly reduced when either the radiative feedback or the 
magnetic field are present. However when both are treated 
simultaneously, the fragmentation is very significantly 
reduced because magnetic field focusses the gas which 
leads to a more efficient radiative feedback and higher
temperatures.

\end{document}